\documentclass[twocolumn,aps,jcp,preprintnumbers,yfonts,eufrak,amssymb,amsfonts]{revtex4}

\newcommand{\be}{\begin{equation}}
\newcommand{\ee}{\end{equation}}
\newcommand{\bea}{\begin{eqnarray}}
\newcommand{\eea}{\end{eqnarray}}
\newcommand{\bd}{\begin{displaymath}}
\newcommand{\ed}{\end{displaymath}}

\usepackage{graphicx}
\usepackage{dcolumn}
\usepackage{bm}
\usepackage{mathrsfs}
\usepackage{amsmath}
\usepackage{amssymb}
\usepackage{amsfonts}
\usepackage{wasysym}
\usepackage{mathbbol}
\usepackage{calrsfs}
\DeclareMathAlphabet{\mathpzc}{OT1}{pzc}{m}{it}

\begin{document}

\preprint{INT/MolEl-p7}

\title{Pride, Prejudice, and Penury of {\it ab initio} transport calculations for
  single molecules}

\author{F. Evers$^{1,2}$}
\author{K. Burke$^{3}$}%
\affiliation{$\mbox{}^1$ Institut f\"ur Theorie der Kondensierten Materie,
 Universit\"at Karlsruhe, 76128 Karlsruhe, Germany}
\affiliation{$\mbox{}^2$ Institut f\"ur  Nanotechnologie,
  Forschungszentrum  Karlsruhe, 76021 Karlsruhe, Germany}
  \affiliation{$\mbox{}^3$ Department of Chemistry, University of
    California, Irvine, California 92697, USA}


\date{\today}

\begin{abstract}
Recent progress in measuring the transport properties of
individual molecules has triggered a substantial demand for
{\it ab initio} transport calculations. 
Even though program packages are commercially available and placed on  
custom tailored to address this task, reliable information often is
difficult and very time consuming to attain in the vast majority of 
cases, namely when the
molecular conductance is much smaller than $e^2/h$.
The article recapitulates procedures for molecular transport
calculations from the point of view of
time-dependent density functional theory.
Emphasis is describing the foundations of
the ``standard method''. Pitfalls will be uncovered
and the domain of applicability discussed. 
\end{abstract}

\maketitle

\section{Introduction}

In an impressive
sequence of experiments, it has recently been demonstrated that
measuring the current voltage (I-V) characteristics of an individual
molecule has become
feasible. \cite{reichert02,kubatkin03,djukic:161402,li05,loertscher06,jo06}
Each
molecule is an interesting species in itself exhibiting individual
signatures in each IV-curve,
such as step positions and heights \cite{elbing05} or
inelastic excitation energies. \cite{heersche:206801}
For this reason, a clear demand for {\it ab initio} transport
calculations of single molecules has emerged in recent years. 

Such calculations are a difficult enterprise, because they
must meet simultaneously two
requirements. Powerful methods exist to deal with each one 
separately, but the combined problem still is one of the 
challenging adventures of theoretical physics and quantum chemistry.

Difficulty number one is that a molecule is a genuine many-body
system, where the mutual interaction of the particles is important for
understanding its properties. These include, in particular, the energy
and shape of the (effective) molecular quasiparticle orbitals.
Some of the salient aspects, such as the position of the molecule's
atoms, the symmetry of molecular orbitals and their relative energies,
are often described accurately by effective single-particle theories,
such as density functional theory. For less basic questions,
concerning for instance excitation energies or details of the
electronic charge distribution, polarization, and charging effects, an
advanced machinery equipped with methods and codes from quantum
chemistry and electronic structure calculations is available.

Difficulty number two is related to the fact
that a transport calculation investigates
the effect of coupling the molecule to a macroscopic electrode,
i.e. a reservoir with which particles (and energy) can be exchanged.  
It is the associated broadening of the molecular energy levels which
is supposed to be understood quantitatively in transport calculations
and therefore the coupling
has to be modelled with great detail and care.

While the first difficulty can be resolved for sufficiently small
molecular systems, the second one requires including many
electrode atoms, i.e., a great number of degrees of freedom,
in order to properly extrapolate to the macroscopic
limit. These conditions are mutually exclusive (almost), and this is the
particular challenge in molecular scale transport calculations with
{\it ab initio} methods. 

In reality, any {\it ab initio} transport calculation begins 
with a compromise accepting strong, often uncontrolled, approximations when
dealing with one of the two mentioned difficulties. In the present
``standard approach to molecular conductance''
\cite{brandbyge02,xue01,paulsson02b} a drastic simplification on
the many-body side is being made. One is accepting the Kohn-Sham energies and orbitals
that appear in structure calculations based on ground-state density functional theory
(DFT) as the legitimate single particle states for a selfconsistent
scattering  theory of transport. The procedure has the tremendous advantage
that including the reservoirs is then a very well defined process, if
sufficient care is taken. For practical purposes, a formulation in terms
of non-equilibrium Green's (or ``Keldysh'') functions is advantageous and
therefore often used. \cite{kadanoff62} For non-interacting particles 
the theory is equivalent to a Landauer-B\"uttiker theory of
transport\cite{landauer57,buettiker86}. For KS-particles a slight
generalization is introduced, in which the electronic charge
distribution is calculated selfconsistently in the presence of the bias voltage. 

Clearly, the use of DFT for the scattering states
already includes many non-trivial interaction effects
beyond the electrostatic Hartree interaction -- Fermi
liquid (FL) renormalizations in the language of condensed matter physics -- even on
the level of local or gradient corrected density approximations
(LDA,GGA). \cite{dreizler90}
Still, ground and excited states of correlated electron systems are {\em not} single
Slater determinants and therefore the validity of applying scattering theories
designed for non-interacting particles to interacting systems is not
straightforward to establish. 
Moreover, including correlation effects beyond FL renormalizations,
which can be very important for transport characteristics
like the Coulomb blockade or the Kondo resonance, 
 is not inconceivable in a single determinant theory, 
 but it certainly requires 
density functionals advanced far beyond LDA.
In this respect it is tempting to use the advanced machinery of quantum
chemistry to calculate a better approximation for correlated many-body
states. Including more than one Slater determinant, however, 
this approach also has a serious drawback because it is limited to relatively small
system sizes. The bare molecule appearing in typical transport experiments
consists of typically 5-10 aromatic rings, which is a size already at the
limits of what correlated methods could still reasonably deal with. Including
in addition 10-200 metal atoms in order to accurately model the coupling
to the leads in a controlled way appears to be out of reach at present. 
For this reason, only very few attempts limited to small molecules have been
made in this direction. \cite{delaney:036805}

In this article, we describe three principle approaches to transport
calculations based on time-dependent density functional theory (TDDFT). 
We shall first present a brief account of the basic principle strategies.
Then, we explain in more detail one of them, the standard method of
{\it ab initio} transport calculations. In Section \ref{sII},
we discuss an attempt to justify the procedure from
the point of view of TDDFT \cite{stefanucci:195318,stefanucci04,evers03prb},
list loose ends and apparent conceptual difficulties.

Since the exchange-correlation potential $V_{\rm XC}({\bf x})$ is not known
exactly, in any practial calculation approximations like LDA 
have to be admitted. These are not controlled any more when 
one deals with realistic molecules. As a consequence, 
in addition to conceptual problems, appreciable artefacts related to
approximate functionals can emerge, which have been well studied for
standard DFT applications in quantum chemistry and electronic structure
theory, and which carry over to transport calculations as well. \cite{reimers03}
A brief list of deficencies most important for transport purposes has
been included in Section \ref{sIII}.

\section{\label{sII} TDDFT and transport}

Time dependent DFT  is a well established
generalization of (ground state) density functional
theory and has been introduced by Runge and Gross \cite{runge84}
and expanded on by van Leeuwen \cite{leeuwen99}.

\begin{description}
\item[RG-Theorem:] For any interacting fermion system
there is a unique dual system of {\it non}-interacting fermionic quasiparticles 
with the following property: the time dependent density of original and dual particles is
identical for any driving field $V_{\rm ex}(t)$;
the time evolution of dual (``Kohn Sham'' or KS) 
fermions is governed by a Schr\"odinger-type equation
decorated with a Hartree term and an exchange correlation
potential $V_{\rm XC}[n]$, which can be expressed as a functional of 
the time dependent particle density and its history, $n({\bf x},t)$ .
\end{description}

Because the Runge-Gross-Theorem guarantees that the dual system
delivers the exact time evolution of
the interacting particle density, also longitundinal transport currents can be
calculated by exploiting the continuity equation \cite{VT04}, 
$$
\dot n({\bf x},t) + \nabla\cdot {\bf j}({\bf x},t) = 0,
$$
where a dot denotes a time-derivative.
This observation underlies all applications of TDDFT to transport.

Quite generally, transport can be investigated in several different languages, which all
are  equivalent in the regime where their validity overlap. Even
though we're ultimately interested in the standard method, the TDDFT
version of the others will give valuable information, too. 
Therefore we shall briefly discuss them as well. We begin, however, by  
recalling the basic formalism of TDDFT.

\subsection{\label{sIIA} TDDFT formalism}

TDDFT is a machinery for propagating a density in
time, not a many body wavefunction. 
Hence, as a prerequisite for applying the method
an initial density ($t{=}0$) is required. It needs to be represented 
as a single Slater determinant  $|0\rangle$  
constructed from a (complete) set of effective 
single particle states $\phi_m$ . 
This is always possible, if at $t{<}0$
the system is in its ground state; then the KS-orbitals of ground state DFT are
obvious candidates for $\phi_m$.
In this case, one has for the density matrix at $t=t'=0$ 
\be
   n({\bf x},{\bf x'}) = \sum_{m}^{\rm occ.} \phi^*_{m}({\bf
     x})\phi_m({\bf x'}).
   \label{e1}
\ee
Time evolution of the state $|0\rangle$ together with its density matrix 
is mediated via $H_{\rm s}$:
\be
H_{\rm s} = {-}\frac{1}{2m} \! \int \! d{\bf x}\ \psi^\dagger({\bf x})\Delta \psi({\bf x})
  + V_{\rm H}[n(t)] + V_{\rm XC}[n(t)] + V_{\rm ex}(t)
\label{e2}
\ee
with 
\be
V_{\rm H}[n(t)] = \frac{1}{2} \int \!d{\bf x}\ v_{\rm H}[n]({\bf x},t)
\ \psi^\dagger({\bf x})\psi({\bf x})
\ee
where $v_{\rm H}({\bf x},t) = \int \! d{\bf x'}\ n({\bf x'};t)/|{\bf x}-{\bf x'}|$ and 
\be
V_{\rm XC}[n(t)] =  \int \!d{\bf x}\ d{\bf x'} \ v_{\rm xc}[n]({\bf
  x},t) \ \psi^\dagger({\bf x})\psi({\bf x}).
\ee
The nontrivial aspects originate from the fact that the orbitals 
$\phi_m(t)$ are not eigenstates of $H_{\rm s}$ at $t>0$. 

$H_{\rm s}$ can be explicitly time dependent 
in the probing potential, $V_{\rm ex}(t)$. An implicit dynamics exists 
via the Hartree and exchange-correlation terms, that depend on the time dependent particle
density $n({\bf x},t)$. A few remarks about the potential 
$v_{\rm xc}[n]$ debuting here are in place. 

(a) The exchange correlation potential $v_{\rm xc}[n]$ is not
just a density functional.
Its precise definition
requires specification of the initial many body state at $t=0$. 
But, for an initial non-degenerate ground state the dependence on the
initial wavefunction is replaced by the initial density, thank to the
Hohenberg-Kohn theorem.

(b)
TDDFT strictly applies only to {\it finite} systems, and a
generalization that uses the current as the basic variable is needed
for infinite systems. 

(c) In practice, it might be advantageous for the construction of useful approximations
to allow for a more general, offdiagonal structure
of $v_{\rm xc}[n]({\bf x},{\bf x'};t,t')$
that could also include (time dependent) gauge fields. 
Similarly, one can also consider  
$v_{\rm xc}[n]$ as a functional of the full density operator
$n({\bf x},{\bf x'};t,t')$, rather than only its diagonal elements, which is the
particle density. 
Thus additional observables, like the current density, are introduced
into the Hamiltonian. In statistical mechanics this is a 
standard recipe in order to eliminate a history dependence in
kinetic equations\cite{brenig89}, and here it serves exactly the same
purpose \cite{vignale:2037}. 

\subsubsection*{\label{sIIA1} Structure of XC potential if
  $V_{\rm ex}$ is weak}

In this subsection, we analyze the non-equilibrium piece of the $v_{\rm xc}[n]$
potential. The idea is to exploit the fact, that it can be related  to  
known correlation functions if the probing potential  $V_{\rm ex}$ is weak. 

We begin by recalling some basic facts of the theory of
linear response for TDDFT. 
Since the density evolution of 
dual and original system, $n({\bf x},t)$, coincide, 
they exhibit in particular the same susceptibility for the 
density, $\chi({\bf x},{\bf x'},t-t')$, which describes the linear 
response to $V_{\rm ex}(t)$. 
What is usually measured is not the response to a 
probing but rather to the total electric field, 
which is the sum of external
and induced (screening) fields. An important example is the linear
conductance. It is the ratio of the current and the
measured (total) electrostatic 
voltage drop at the resistor: $I/V_{\rm bias}$. 

The corresponding response function is the 
Hartree-irreducible correlator: 
$$
    \chi_{\rm irr}^{-1} \equiv \chi^{-1} + f_{\rm H}
$$
($f_{\rm H}({\bf x},{\bf x'}){=}1/|{\bf x}{-}{\bf x'}|$).
In TDDFT the operator $\chi_{\rm irr}$ can be decomposed even further.
Namely, the TDDFT Hamiltonian $H_{\rm s}$ has two pieces
that react to density modifications. 
In addition to $V_{\rm H}$ incorporating the electrostatic screening, there is
also an induced effect on $V_{\rm XC}$,
\be
    \chi^{-1}_{\rm irr} = \chi^{-1}_{\rm KS} - f_{\rm xc},
\label{e5}
\ee
which can be split off in the same manner as $f_{\rm H}$. 
The truncated correlator $\chi_{\rm KS}$ describes the (bare) response of 
ground state DFT. 
The promised connection between the correlator $\chi_{\rm irr}$ and 
$v_{\rm xc}[n]$ is mediated via the XC kernel:
\be
f_{\rm xc}({\bf x},{\bf x'};t-t'){=}\delta v_{\rm xc}[n]({\bf x},t)/\delta n({\bf x'},t').
\label{e6}
\ee
Relations (\ref{e5},\ref{e6}) are very useful, 
because due to a beautiful series of works by Kohn, Vignale and
collaborators there is a simple approximation to the  non-equilibrium
contribution to $\chi_{\rm irr}$. \cite{vignale:2037,vignale97}
In fact, these authors reveal the full hydrodynamic structure of $\chi^{-1}_{\rm irr}$
by exploiting the relation to the phenomenological theory of quantum liquids. 

In their analysis, $f_{\rm xc}$ is the sum of two very 
different pieces:
\be
f_{\rm xc} = f_{\rm xc}^{\rm adia} + f_{\rm xc}^{\rm non-eq}
\ee 
where $f_{\rm xc}^{\rm adia}{=}\delta v_{\rm xc}^{gs}[n]/\delta n$ and
a gradient expansion of $_{\rm xc}^{\rm non-eq}$ is given by $f_{\rm xc}^{\rm VK}$.
The first term is fully analogous to $f_{\rm H}$ and it 
describes the exchange-correlation screening 
of the ground state DFT Hamiltonian to the probing field. 
Only in the second piece many body effects of a genuinely
non-equilibrium nature appear: 
it incorporates the visco-elastic response of the electronic quantum liquid. 
Since the emerging term is local and dissipative, the corresponding forces are not
conservative. This means, that $f_{\rm VK}$ cannot be incorporated as a 
pure density coupling -- i.e. a potential term -- in $H_{\rm s}$, 
but gives rise to a (time dependent) gauge field, instead, and is only
described as a current-dependent kernel.

\subsection{External driving field}

In  this section we return to the transport problem in TDDFT.
To investigate transport currents generated by an external driving field,
one needs to supplement $H_{\rm s}$ with an inhomogenous electric 
potential, $V_{\rm ext}(q,\omega)$, together with electrodes.
\footnote{Equivalently, a time dependent vector
  potential may also be introduced.}
{\em Practical} applications with this approach in TDDFT
suffer from the fact, that before one can perform the
$dc$-limit ($\omega{\to}0$), one first has to take the 
thermodynamic limit ($q{\to}0$) of infinite system size. In realistic 
calculations this is usually a very cumbersome excercise. 

From a {\em conceptual} point of view, thinking about currents as generated by
weak external driving fields is rewarding, however, because one is led back
to the Kubo formula and the theory of linear response. 
We thus can directly apply results from the preceeding section 
and in particular investigate the effect of visco-elasticity on the current
flow.

\subsubsection{\label{sIIB1} Kubo formula}

The Kubo formula provides an exact relation between 
the current density and the driving  
external and induced (effective) electric fields, 
that appear in the TDDFT calculation \cite{koentopp:121403}:
\be
j({\bf x},\omega) = \int \! d{\bf x'} 
\ \sigma_{\rm KS}({\bf x},{\bf x'},\omega)\left[ 
E_{\rm ex} + E_{\rm H} + E_{\rm xc} \right]({\bf x'},\omega).
\label{e8}
\ee
where $\sigma$ is the non-local conductance tensor and where the
electric fields derive from the potential terms given in Eq. (\ref{e2}). 
As always, susceptibilities for particle densities $\chi_{\rm KS}$ and 
currents $\sigma_{\rm KS}$ are related 
via the continuity equation:
\be
    \partial_{t} \chi_{\rm KS}({\bf x},{\bf x'},t) = - \sum_{i,j=1}^{3}
\nabla_{i}\nabla_{j}\sigma_{{\rm KS},ij}({\bf x},{\bf x'},t).
\ee
The structure analysis \ref{sIIA1} suggests to split the full current density 
into a bare response, $j_{0}$, and a remaining piece, $j_{\rm VK}$:
\be
 j = j_{0} + j_{\rm VK}.
\label{e9}
\ee
We discuss the second term first. It is driven by a force
field, $E_{\rm VK}$,  
\be
    j_{{\rm VK}}({\bf x},\omega) = \int \! d{\bf x'} 
\sigma_{\rm KS}({\bf x},{\bf x'},\omega) E_{{\rm VK}}({\bf x'},\omega).
\ee
that is associated with $f_{\rm VK}$. In the hydrodynamic limit considered by
Vignale et al. \cite{vignale97} it has an
interpretation as visco-elastic force internal to electron liquid, that can be 
described by the stress tensor, $\varsigma_{ij}$
(unperturbed density: $n_{0}({\bf x})$): 
\be
    E_{{\rm VK},i}({\bf x},\omega) =  n_0({\bf x'})^{-1}\sum_{k{=}1}^3\ \nabla_k
    \varsigma_{ik}({\bf x'},\omega).
\label{e11}
\ee

The relative magnitude of $j_{\rm VK}$ is small as compared to
$j_{0}$,
since the viscosity, $\eta$, of the electron liquid is quite small. Based on 
the homogenous value of $\eta$ in two and three spatial dimensions
a rough analytical consideration can 
give an estimate. \cite{koentopp:121403}
The ratio $j_{\rm VK}/j_{0}$ 
is expected to be typically of the order of 10\% or less. 
In a numerical study the viscous corrections to the conductance of the benzene-dithiol molecule
have been explicitly calculated.
The effect is small, roughly 5\%, as expected. \cite{sai05}

Note, however that the previous conclusion is to be taken with a grain of
salt.  Strictly speaking, the explicit derivation of  
Eq. (\ref{e11}) assumes, that the inhomogeneities, which provide the
``surface'' for the viscous friction to appear,  
are very smooth: in the period $\omega^{-1}$ of the probing field, 
the electron should travel a distance not larger than the
typical spatial scale $\ell$ on which the inhomogenuous
background changes, $v_{\rm F}/\omega{\ll} \ell$.
Applying this condition to molecules and assuming $k_{\rm F}\ell \sim 1$, one would
get $v_{\rm F}k_{\rm F} \ll \ell \omega$, which is satisfied only 
at optical frequencies of the order of eV, but not in the {\it dc} limit.
Nevertheless, the qualitative finding of the mentioned 
estimates 
-- namely that viscocity effects tend to be small --
should be indicative, because the
(transverse) momentum exchange between electrons at low temperatures is a 
rare process due to phase space constraints. We believe that 
this basic principle pertains to electrons in a molecule as well. 
\footnote{An additional caveat should be mentioned, also: the
  viscosity $\eta$ not only has a real (dissipative) but also an
  imaginary (reactive) piece. The effect of the latter
  has not been investigated so far and we have ignored it, here.}

We thus propose that the dominating contribution in  Eq. (\ref{e9})
is given by the response to the reactive forces,
\be
j_{0}({\bf x}) = \int \! d{\bf x'} 
\ \sigma_{\rm KS}({\bf x},{\bf x'})\left[ 
E_{\rm ex} + E_{\rm H} -\nabla_{\bf x'}v_{\rm xc}^{\rm gs} \right]({\bf x'}),
\label{e13}
\ee
where our notation suppresses the $\omega$-dependence.

\subsubsection{\label{sIIB2} Bias voltage and Kohn-Sham voltage drop}

The expression (\ref{e13}) for the current density can be simplified,
if we assume, that the dependence of the forces on the
coordinate, ${\bf x}_\perp$, perpendicular to the current path, $z$, is negligibly
weak. This assumption is not necessarily a good one for quantitative questions,
but it allows us to discuss more clearly the difference between 
$\sigma_{\rm KS}$ and the conductivity, $\sigma_{\rm irr}$, measured in typical
transport experiments:
\be
j({\bf x}) = \int \! d{\bf x'} 
\ \sigma_{\rm irr}({\bf x},{\bf x'})\left[ 
E_{\rm ex} + E_{\rm H}\right]({\bf x'}).
\label{e14}
\ee

If we make the proposed step and neglect the ${\bf x}_\perp$
dependence of the forces, we can integrate both sides of (\ref{e14}) 
over any cross section,  and obtain
\be
    I = G \ V_{\rm bias} \label{e20}
\ee
with a conductance 
\be
    G = \int \! d{\bf x}_\perp d{\bf x'}_\perp 
    \sigma_{\rm irr}({\bf x}_\perp,{\bf x'}_\perp;z,z').
\label{e15}
\ee
Due to particle number conservation, the cross-sectional integrals render the
sum independent of $z,z'$ in the {\it dc}-limit, $\omega{\rightarrow}0$. 
As usual, the bias voltage is given by 
\be
V_{\rm bias} = \int_{{\bf x}_l}^{{\bf x}_r} \! d{\bf s} \ \left[
E_{\rm ex} +
  E_{\rm H}[n]\right]({\bf s}).
\label{e18}
\ee
where ${\bf s}(t)$ is any path connecting the left   
with the right hand side of the molecule. $V_{\rm bias}$ should be picked up 
between points ${\bf x}_l$ and ${\bf x}_r$ sufficiently far
away from the scattering region, in the near asymptotics, where the
electrostatic potential energy surface has turned constant.

The same procedure can be repeated also for Eq. (\ref{e13})
\be
    I = G_{\rm KS}\ V_{\rm KS} 
\label{e18b}
\ee
with an expression for $G_{\rm KS}$ completely analogous to (\ref{e15}). 
We have introduced the KS-voltage drop, $V_{\rm KS}$,
given by the
sum  of {\em all} KS-forces, Eq. (\ref{e13}), along ${\bf s}(t)$, 
\be
V_{\rm KS} = V_{\rm bias} + V_{\rm xc}; 
\qquad V_{\rm xc} = -\int_{{\bf x}_l}^{{\bf x}_r} \! d{\bf s} \ 
E_{\rm xc}({\bf s}).
\label{e19}
\ee
In the spirit of Eq. (\ref{e9}), we can decompose the deviation,
$V_{\rm xc}$, of the measured bias and $V_{\rm KS}$ into two pieces
\be
   V_{\rm xc} = \int_{{\bf x}_l}^{{\bf x}_r} \! d{\bf s}
   \ \nabla_{\bf s} v^{\rm gs}_{\rm xc}[n] ({\bf s})
- \int_{{\bf x}_l}^{{\bf x}_r} \! d{\bf s} \ 
E_{\rm VK}({\bf s}).
\label{e20}
\ee
The first term can be integrated trivially:
$v_{\rm xc}^{\rm gs}[n]({\bf x}_l)-v_{\rm xc}^{\rm gs}[n]({\bf
  x}_r)$. If the left and right hand side electrodes consist of the
same material, then this difference can be non-vanishing only due to
long range terms in  $v^{\rm gs}_{\rm KS}$. In particular,
local approximations like LDA or GGA cannot give a finite contribution
in the XC part. 
The second term in Eq. ({\ref{e20}}) describes the genuine
non-equilbrium forces, that result from the viscosity of the elctron
liquid discussed above. 

Eqs. (\ref{e18b}) and (\ref{e19}) demonstrate, that a KS-particle
behaves under bias very differently from a physical quasiparticle.
The only long range forces, that the physical particle realizes upon
applying $V_{\rm ex}$ are of the pure Coulomb type.
For this reason, the bias voltage must be exactly
equal to the difference in electro-chemical potentials:
$$
    V_{\rm bias} = \mu_\mathcal{L} -\mu_\mathcal{R}.
$$
Interaction terms (beyond Hartree-level) do not occur. By contrast,
the KS-particle experiences an effective voltage $V_{\rm KS}$, that
can be quite different from $V_{\rm bias}$. This way of including
interaction effects by adding the corrective term $V_{\rm xc}$
to the voltage drop is not very physical. A difficulty appears, that
comes back on us in the next two subsections.

\subsection{\label{sIIC} Initial value problem}

The second access to transport investigates an initial value 
(so called ``relaxation'')
problem without any reference to a driving external field $V_{\rm ex}$. 
One considers a molecule and two reservoirs, left and right
($\mathcal{L,R}$). At $t<0$ the molecule is coupled to and in
equilibrium with $\mathcal{L}$. At $t=0$ a coupling to $\mathcal{R}$
is being switched on and the time evolution with $H_{\rm s}$ begins, as described in 
section \ref{sIIA}. 
A current, $I$, starts to flow at $t>0$,
if $\mathcal{L}$ and $\mathcal{R}$ are not in equilibrium with one another. 
$I$ is related to the time derivative of the number of particles in the electrodes, 
$N_\mathcal{L,R}$, 
\be
I(t) = -\dot N_\mathcal{L} = \dot N_\mathcal{R}.
\label{e21}
\ee
As in the previous case, 
the current may be obtained via the continuity equation 
from an explicit TDDFT propagation of the
electronic density. Since relaxation involves processes on all time scales,
in principle the response function $\chi(\omega)$  can be extracted
at all frequencies larger than the inverse
observation time. 

Also, the drawbacks of this approach are similar to the previous case: 
one has to include big reservoirs and to propagate many  time steps  
if long time, low frequency properties, 
such as steady-state currents, are to be addressed. 
Due to this difficulty, the relaxation method with TDDFT has been applied mainly 
to obtain the high frequency response.
In an incarnation where a step potential is switched on at $t{=}0$, 
it has also been used for transport
studies in non-interacting model systems, but not for a full-fledged 
realistic TDDFT calculation, yet. \cite{kurth:035308}
However, we mention that  an application of the
method for model studies of strongly correlated transport
in interacting 
Hubbard chains has recently been very successful
using the density renormalization group method.  \cite{bohr05}

For our purpose, the formulation of transport in terms of an 
initial value problem is conceptually 
important, because it 
allows us to link time propagation of the density 
(TDDFT) with DFT-scattering theory. 

Indeed, let us perform the following
``Gedanken experiment'' in which we allow ourselves to
work with perfect reservoirs, and where we can
do the time propagation of the density up to any
time we wish. At the initial stages, $0<t\ll t_{\rm trans}$,
we shall encounter transient
phenomena, which render the particle density near the contact region
time dependent. Only at a much later stage, $t\gg t_{\rm trans}$
we arrive at the asymptotic non-equilibrium situation. 

In order for the usual scattering formalism to be applicable, 
the asymptotic current carrying state of TDDFT, $|{\rm QS}\rangle$, should meet the following conditions: 
\begin{description}
\item[c1:] At zero temperature $|{\rm QS}\rangle$ is a single
  Slater determinant of left and right moving scattering states,
  $\psi_{l,r}$, which
  are eigenstates of the asymptotic TDDFT Hamiltonian $H_{\rm qs}$.
  The associated quasi-static KS-density matrix
  (definition see Eq. (\ref{e22})) is invariant under
  time translations: $n_{\rm qs}({\bf x},{\bf x'};t-t')$. 
\item[c2:]
  The potential $v_{\rm xc}[n]$ takes an asymptotic
  form which is independent of the history.  
\item[c3:] The KS-scattering states are
  occupied according to Fermi-Dirac distributions, $f_\mathcal{L,R}$,
  carrying the temperature and chemical potential of
  the reservoirs that they emanate from. 
\end{description}

\subparagraph{Discussion:}
Very little rigorous is known about the true nature of the
non-equilibrium state of the interacting electron liquid.  
For this reason our discussion can be no other but very qualitative.

Condition {\bf c1} puts a strong requirement on the physical relaxation
process. Because the equal time density matrix
$n_{\rm qs}({\bf x},{\bf x'};0)$ is time independent
at $t\gg t_{\rm diss}$,
particle and current densities must have become stationary:
\begin{description}
\item[R:] After transient phenomena have died out,  at $t\gg t_{\rm
  trans}$ a quasi-stationary non-equilibrium state is reached. 
``Quasi-stationarity'' in this context is meant in the strong sense, 
in which the time evolution of the particle and current densities have
come to a standstill. 
\end{description}
It is plausible, that this requirement is always fullfilled 
in the linear regime of small voltages
$\mu_\mathcal{L}{-}\mu_\mathcal{R}$.
\footnote{For a thorough discussion of a related problem,
  see Chapt. 3 of Ref. \onlinecite{brenig89}.}
For non-interacting particles this is certainly true also in the
non-linear case. The situation is much less clear for interacting
particles in the non-linear voltage range. In fact, due to the
non-linear nature of the kinetic equations one suspects that phenomena
like turbulence should occur.  \cite{agosta05}
This would imply densities and currents
fluctuating in time even at $t\gg t_{\rm trans}$,
so that {\bf R} is not strictly satisfied.

If indeed the quantum liquid goes turbulent, then part of the
memory of the intial conditions never is lost. This is, because 
even small microscopic details in the initial values of the relevant kinetic fields
(particle densities, currents etc.) will in general invoke a 
different dynamics at later times. For this reason, validity of
{\bf c2} is not guaranteed.
However, usually there is no interest in
a precise set of initial conditions. Possibly, a suitable averaging procedure
(either over initial conditions or over a small time interval) could
reestablish {\bf c2} in an effective sense.

Returning to the case of small biases, let us emphasize that even if
$|QS\rangle$ is a single Slater determinant, validity of ${\bf c3}$ is
not automatically guaranteed.
In order to see this,
we imagine a surplus of particles in one reservoir ($T{=}0$), so
that an electro-chemical potential difference maintains a particle
flow to the other reservoir.  If it is correct, that 
all current is carried by the scattering states in energy window situated
between $\mu_\mathcal{L}$ and $\mu_\mathcal{R}$, then we find that the
current linear in the voltage is necessarily given by
$G_{\rm KS}(\mu_\mathcal{L}{-}\mu_\mathcal{R})$. The correction term
$V_{\rm xc}$ is missing, so we arrive at a statement contradicting
Eq. (\ref{e18b}). This is the difficulty, already alluded to at the end of the
last subsection. We will come back to it again, also in the following
one. 

\subsection{\label{sIID} Scattering approach}

It is {\it the} advantage of scattering theory that all information 
is encoded in scattering states and no reference to time propagation is
being made. The idea is to replace the initial value problem 
by an equivalent boundary problem. This is
the philosophy adopted by the standard method of
molecular transport calculations. 
Its persuasive, charming aspects are: (i) the method is stationary
and (ii) the reservoirs are relatively easy to include. 

The most important problematic aspect is that up to now
a rigorous justification of the
approach has not been given, and it is not obvious that it should exist. 
The validity of the scattering formalism has been rigorously
established only for non-interacting particles. 
Whether the conditions {\bf c1}-{\bf c3} formulated in the preceeding section are
really met, so that the treatment is legitimate also
for KS-particles, is not fully clear at present. If they are taken for
granted, the following selfconsistency procedure can be justified,
which in essence is the standard method.  

One starts with a guess for the equal time density operator
$n_{\rm qs}({\bf x},{\bf x'})$. Condition {\bf c2} ensures, that nothing more is
needed in order to construct a first approximation for the
Hamiltonian $H_{\rm qs}$ -- provided the functional $v_{\rm xc}[n]$ is
given, of course. 

In order to start the next iteration, one should know how to construct
a better guess for the density operator from $H_{\rm qs}$. 
This is where {\bf c3} and again {\bf c1} kick in: according to {\bf c1} scattering
states can be found as the eigenstates of $H_{\rm qs}$, which then 
can be filled up successively in order to obtain $|QS\rangle$ and
the improved density
matrix. The procedure is completely analogous to the case of
non-interacting particles: 
\bea
n_{\rm qs}({\bf x},{\bf x'}) &=& \sum_{l} \ f_\mathcal{R}(\epsilon_l) 
\ \psi_l^*({\bf x}) \psi_l({\bf x'}) \nonumber\\ && + 
\sum_{r} \ f_\mathcal{L}(\epsilon_r) \ \psi_r^*({\bf x})\psi_r({\bf x'}).
\label{e22}
\eea
At the end of the iteration cycle, selfconsistency is reached.
This means that scattering states have been found in an effective
potential that incorporates already the shifts in the charge
distribution characteristic of the non-equilibrium situation which
also generates the current. 

The bottom line is that 
under the assumptions {\bf c1}-{\bf c3} there are only two minor
differences between a ground state calculation and a
standard {\it dc}-transport calculation:
\begin{enumerate}
\item A non-equilibrium density operator
calculated from Eq. (\ref{e22}) replaces the ground state expression
Eq. (\ref{e1})
(to be obtained with $f_\mathcal{L}=f_\mathcal{R}$).
\item In principle, a quasi-stationary functional should replace the
ground state functional. In actuality, due to lack of any better
choice, common ground state functionals are employed leading to
additional artefacts, Section \ref{sIII}.  
\end{enumerate}


\subsection{Standard method and Kubo formula: discrepancies}

The calculation of the {\it dc}-current can proceed
directly from Eq. (\ref{e22}) which leads to a Landauer-Buttiker type
description,
\be
I = \int \! dE \ T(E,\mu_\mathcal{L},\mu_\mathcal{R}) \left[
  f_\mathcal{L} - f_\mathcal{R} \right],
\label{e23}
\ee
where the transmission function, $T$, has been introduced.
The kernel $T(E,\mu_\mathcal{L},\mu_\mathcal{R})$ can be expressed by
the resolvent operator
$$
G(E) = (E-H_{\rm qs}-\Sigma_\mathcal{L}-\Sigma_\mathcal{R})^{-1}
$$
and self energies $\Sigma_\mathcal{L,R}$, which
represent the boundary
conditions at the surface of the electrodes. \cite{evers05book}
One has
\be
    T(E,\mu_\mathcal{L},\mu_\mathcal{R}) = \text{tr} \ \Gamma_\mathcal{L}
    G \Gamma_\mathcal{R} G^\dagger
    \label{e24}
\ee
where $\Gamma_\mathcal{L,R}=i(\Sigma-\Sigma^\dagger)_\mathcal{L,R}$. \cite{evers03prb}
For non-interacting particles, Eqs. (\ref{e23},\ref{e24}) are equivalent to the
Landauer formula and give the exact current. \cite{brandbyge02}
The Landauer conductance, 
$G_{\rm qs}{=}T(\epsilon_{\rm F}) e^2/h$, is the
response of the current linear in the applied electro-chemical potential difference
\be
    I = G_{\rm qs} (\mu_\mathcal{L}{-}\mu_\mathcal{R}).
\ee
(We choose the nomenclature in the spirit of Sec. \ref{sIIB2}.) 
For non-interacting particles, rigorous results exist, which show
that the Landauer conductance 
and the conductance obtained from
the Kubo-conductivity, $\sigma_{\rm irr}$  coincide. \cite{baranger89}
This is true, because in the absence of interactions 
\be
G=G_{\rm KS}=G_{\rm qs}.
\ee

However, the KS-particles are {\em not} truely non-interacting. This expresses itself 
in the fact that $G{\neq} G_{\rm KS}$, in general,
because $V_{\rm XC}{\neq}0$.
In the present standard approach, one has $G_{\rm qs}{=}G_{\rm KS}$.
Therefore, a term in the current proportional to $V_{\rm XC}$ is
ignored. In order to include this term, one would have to manipulate
the XC-functional used in $H_{\rm qs}$ such that
the {\em bare} {\it dc}-current response becomes $\chi_{\rm irr}$
rather than $\chi_{\rm KS}$. Note, that it is known that even within
LDA the terms ignored can be important. Including them on the level
of ``adiabatic'' LDA can shift resonances and therefore have a
substantial impact on a transmission characteristics.
\cite{gonze99}

\section{\label{sIII} Ground state DFT: Artefacts of local density functionals}

In this section, we discuss some well-known limitations of common
density functional approximations, and their implications for transport
calculations. \cite{CKBC07}

By common density functional approximations, we will mean the original
local density approximation of Kohn and Sham\cite{KS65}, the generalized
gradient approximation, employing both the density and its gradient,
eg PBE\cite{PBE96}, and hybrids of GGA with exact exchange, such
as PBE0\cite{PBEb96} and B3LYP\cite{B88,LYP88,Bb93}.

There are a variety of related deficiencies of all these approximations.
The first is that they fail for one-electron systems, in which the 
exchange energy should exactly cancel the Hartree, while the correlation
energy should vanish.   The above density functionals generally fail 
this requirement, and are said to have a {\em self-interaction} error,
meaning that the electron is incorrectly interacting with itself.
\cite{PZ81}

A related difficulty is that the ground-state KS potential in such approximations
is poorly behaved.  For a neutral system, the exact KS potential decays
as $-1/r$ for large distances.  But with these approximations, the potential
decays too rapidly, in fact exponentially with distance, due to the 
local dependence on the density.  (Hybrid functionals do have a fraction of
$-1/r$, but not the right amount.)  This leads to potentials that
are far too shallow overall, and HOMO's (highest occupied molecular 
orbitals) that are insufficiently deep.

The exact KS HOMO can be
proven to be equal to the negative of the ionization potential,
but this is not even roughly true for approximate KS potentials, for
reasons given above.  Thus the charge density in tail regions, i.e. where
the density is low, is inaccurate.  Furthermore, whenever a localized
system is in weak contact with a reservoir, so that the average particle number
on the system can be continuous,  the exact KS potential jumps by a 
(spatially) constant amount whenever the particle number passes through
an integer\cite{PPLB82,PL83}.   This behavior is entirely missed by
the above approximations, which smoothly interpolate between either
side of this discontinuity.

These difficulties are either largely or totally overcome by the
use of orbital-dependent functionals.  The first popular one of
these was SIC-LDA, the self-interaction corrected local density
approximation, as introduced by Perdew and Zunger\cite{PZ81}.  
These days, many codes have been developed to handle orbital-dependent
functionals and to find the corresponding KS potential, via the optimized
potential method (OPM), a.k.a. optimized effective potential (OEP).
Exact exchange potentials have the correct decay, their HOMO's are
close to the negative of the ionization potential, and they jump
discontinuously at integer particle number. \cite{E03}

How does all this affect transport calculations?  There are two 
principal effects, one obvious, the other less so. 

In the first, since transport is often a weak tunnelling process, the
position of the molecular levels relative to the leads greatly
affects the calculated current.  If a molecule is weakly coupled
to the leads, there is every reason to think that standard functional
approximations will make huge errors in the calculation of currents.
The levels will be misaligned not only in the equilibrium situation,
but will respond completely wrongly to the transfer of charge into 
a localized molecular orbital. \cite{TFSB05}

These deficiencies in common ground state functionals may be the reason
why present DFT-calculations fail to correctly reproduce
elementary ground state properties that manifest themselves in the
transport characteristics. The most prominent example is the Coulomb blockade
phenomenon, which usually is not reproduced
quantitatively. \cite{arnold06}
It is less well appreciated, that also the
Kondo-effect belongs to this category of phenomena. 
Its manifestation is an extra resonance in the spectral function
of the molecule at $\epsilon_{\rm F}$, the {\it Abrikosov-Suhl resonance}. 
In principle, this resonance is a ground state property. It affects the total charge
on the molecule and for this reason it should be detectable with DFT,
provided an appropriate (so far unknown) functional is used. 

Furthermore, density functional approximations
will artificially smear out a sharp resonance into a much weaker
peak, spread out sometimes over several eV, between the LUMO of the uncharged
molecule and the HOMO of the charged molecule. \cite{koentopp:121403}
This effect has already been demonstrated in calculations on simple
models \cite{TFSB05}, and has recently been seen in a
full OEP calculation. \cite{ke06}
Of course, for molecules that are chemically bonded to the leads,
there is no region of very low density, and the potential should
be reasonably accurate within the common approximations.  So another
question is:  how big an effect is this in real experiments?

The second effect is more subtle, but could be more important in the
chemically-bonded situation, perhaps.  The standard approximations, being local
in nature, yield no XC correction to the potential drop across the molecule.
In the language of the previous section, $V_{\rm KS}$ is identical
to the real electrostatic potential drop, $V_{\rm bias}$.
But there is no reason that this
should be true in reality, or in a more accurate calculation.  Thus an
exchange calculation should produce a finite effect.  As mentioned above,
a current-dependent approximation (VK) indeed produces a small, but finite,
effect.   Such calculations, performed self-consistently and at finite
bias, need to account for this drop, and correct the Landauer formula
to account for it.  This means that the conductance is {\em not} just 
proportional to the
transmission through the self-consistent KS potential. In one spatial
dimension this means, that the
current must be calculated using the total potential drop, including
the XC contribution, and this must be divided by the electrostatic
potential drop.  In this case, it is hard to see how a simple single-particle
effective potential could produce the exact conductance. \cite{koentopp:121403}

\section*{Conclusion}

The various issues that we have discussed -- validity of a scattering
approach, neglect of $V_{\rm XC}$,
deficiencies of commmon density functional approximations --
raise serious doubts about the accuracy of the 
present standard method for transport calculations.
How quantitatively significant these errors are is only poorly
understood, at the moment. For such an estimate,
time and orbital-dependent calculations are an 
important tool. \cite{kurth:035308}
To gain further insight, it is important to go also beyond (TD)DFT.
Work in this direction is in progress.
Proposals include approaches based on a configuration interaction 
\cite{delaney:036805}, the $GW$-method \cite{thygesen06}, and
the LDA$+U$ formalism \cite{pemmaraju06}. 
All of the proposed directions have their virtues and drawbacks. It
remains to be seen which one of them turns out to be most
suitable to deliver conductances of real, large molecules which
require a controlled handling of electrode effects. 

\section*{Acknowledgements}
FE expresses his gratitude to P. Schmitteckert, P. W\"olfle and
E.~K.~U. Gross for valuable discussions. Also, we thank V. Meded
for useful comments on the manuscript as well as
the Center for Functional Nanostructures of the Deutsche
Forschungsgemeinschaft
at Karlsruhe University for financial support (FE).

\bibliography{../../Bib/extern,../BibME/bibNegf,../BibME/bibTechnology,../BibME/bibOwnMolEl,../BibME/bibDft,../BibME/bibExperiments,../BibME/upMolEl,../BibME/bibMethods,../BibME/bibAuChains,../BibME/bibTDFT,../BibME/books,../../Pap.MolEl/BibME/bibStrongCorrel.bib,../../Pap.MolEl/BibME/bibStrongCorrel.bib}

\end{document}